\documentclass[a4paper,11pt]{article}
\pdfoutput=1 

\usepackage{jinstpub} 

\title{\boldmath Mixed (Regular and Chaotic) Dynamics under the Channeling of the High Energy Positrons in [100] Direction of the Silicon Crystal and Podolskiy-Narimanov Distribution}

\author[a,1]{V.V. Syshchenko,\note{Corresponding author.}}
\author[a]{A.I. Tarnovsky,}
\author[b]{and A.Yu. Isupov}

\affiliation[a]{Belgorod State University,\\Pobedy Street, 85, Belgorod 308015, Russian Federation}
\affiliation[b]{Laboratory of High Energy Physics (LHEP), Joint Institute for Nuclear Research (JINR),\\Dubna 141980, Russian Federation}

\emailAdd{syshch@yandex.ru}

\abstract{The character of motion (regular or chaotic) on the quantum level manifests itself in the statistics of the set of the system's energy levels. The completely regular case generates the sequence of the levels with exponential (Poisson) level spacing distribution; the completely chaotic one --- with Wigner distribution. The most interesting case is the co-existence of the regular and chaotic motion domains in the phase space of the system under consideration. The assumption of independent generation of two level sequencies by one chaotic domain and by all regular domains leads to Berry--Robnik distribution. However, the presence of the chaotic motion domain in the phase space affects the levels produced by the regular domains via the so-called chaos-assisted tunneling (CAT) that leads to Podolskiy--Narimanov distribution of the level spacings. This distribution needs the mean amplitude of the tunnel transition and the relative contribution of the regular domains to the mean level density as the parameters. Using their estimations for the transverse motion of the high energy positrons channeling near [100] direction in the silicon crystal we found that Podolskiy--Narimanov distribution demonstrates the best agreement with the level spacing distribution.}


\arxivnumber{2311.06635} 

\proceeding{XIV International Symposium ``Radiation from Relativistic Electrons in Periodic Structures''
and VIII International Conference ``Electron, Positron, Neutron and X-ray Scattering under External Influences''\\
September 18-22, 2023\\
Tsaghkadzor, Armenia}

\begin{document}
\maketitle
\flushbottom

\section{Introduction}
\label{sec:intro}

When a fast charged particle is incident on a crystal at a small angle to any crystallographic axis densely packed with atoms, it can perform the finite motion in the transverse plane. This motion is known as the axial channeling \cite{AhSh, AhSh2, Ugg}. The particle motion in this case could be described with a good accuracy as the one in the continuous potential of the atomic string. During motion in this potential the longitudinal particle momentum $p_\parallel$ is conserved, so the motion description is reduced to two-dimensional problem of motion in the transversal plane. From the viewpoint of the dynamical systems theory, the channeling  particle's motion could be either regular or chaotic. The quantum chaos theory \cite{9, Stockmann, Reichl, Bolotin.book} predicts qualitative differences for these alternatives. 

The most prominent (as well as simple for the analysis) manifestations of chaos in quantum systems are found in the statistical properties of their energy spectra. The semiclassical mean level density is determined by the volume accessible for the motion in the 4-dimensional phase space \cite{Berry}:
\begin{equation}\label{density.via.Berry.Robnik.general}
\rho (E_\perp) = (2\pi\hbar)^{-2} \int dx dy dp_x dp_y \, \delta \left( E_\perp - H(x,y,p_x , p_y) \right) ,
\end{equation}
where $H(x,y,p_x , p_y)$ is the classical Hamiltonian of the transverse motion of the channeling particle.
The quantum chaos theory predicts (see, e.g., \cite{9, Stockmann, Reichl, Bolotin.book}) that the density fluctuations of the energy levels on the uniform background $\rho (E_\perp)$ are qualitatively different for the cases when corresponding classical motion is regular and chaotic. Hence $\rho (E_\perp)$ is not generally a constant the so-called unfolding procedure is used to produce from the original level set $\{E_\perp\}$ the new one $\{\tilde E_\perp\}$ with the unit mean density $\rho = 1$ \cite{9, Reichl}. Hereafter we will discuss the statistic properties of these renormed (unfolded) level sets.

The quantum chaos theory predicts that the energy levels nearest-neighbor spacing distribution of the regular system obeys the exponential distribution (characteristic to Poisson flow)
\begin{equation}
    \label{Poisson}
   p^{}_P(s) = \exp  (-s) 
\end{equation}
(where $s$ is the distance between consequent energy levels of the unfolded set) while the chaotic system obeys Wigner distribution
\begin{equation}
    \label{Wigner}
   p^{}_W(s) = (\pi s/2) \exp (-\pi s^2/4) ;
\end{equation}
the last one describes well the level spacing statistics for the electrons channeling in $[110]$ direction of the silicon crystal \cite{Pov.2015, NIMB.2016, Pov.2021} when the motion above the saddle point of the two-well potential is almost completely chaotic \cite{AhSh2}.

More complex and interesting is the case of co-existence of the domains of regular and chaotic dynamics in the system's phase space. This situation takes the place when the electron or positron is channeling near $[100]$ direction of silicon crystal. The first approach to description of this situation was made by Berry and Robnik \cite{Berry} and, independently, Bogomol'nyi \cite{Bogomolnyi}. These authors have assumed that the spectrum of such mixed system consists of two independent level sequencies: one is generated by all regular motion domains while another --- by (the only) chaotic domain, with the relative densities $\rho^{}_1$ and $\rho^{}_2$ ($\rho^{}_1 + \rho^{}_2 = 1$) respectively. This leads to the level spacing distribution with the only parameter $\rho^{}_1$:
\begin{equation}\label{Berry.Rob}
p^{}_{BR}(s) = \exp \left( -\rho^{}_1 s \right)
\left\{ \rho_1^2 \, \mathrm{erfc} \left( \frac{\sqrt{\pi}}{2} \rho^{}_2 s \right) + \left( 2\rho^{}_1 \rho^{}_2 + \frac{\pi}{2} \rho_2^3 s \right) \exp \left( -\frac{\pi}{4} \rho_2^2 s^2 \right) \right\},
\end{equation}
where 
\begin{equation*}
\mathrm{erfc}\, (x) = \frac{2}{\sqrt{\pi}} \int_x^\infty e^{-t^2} dt = 1 - \mathrm{erf}\, (x) \,.
\end{equation*}

However, the levels in the regular and chaotic sequences are not independent: the interaction between levels changes the level spacing statistics in a way demonstrated by Podolskyi and Narimanov in the series of papers \cite{Narimanov.1, Narimanov.2, Narimanov.3}. They stated that the presence of the chaotic motion domain can increase the tunneling rate between the dynamically isolated regular domains in the phase space (the so-called chaos-assisted tunneling, CAT \cite{Bolotin.book, Narimanov.1, Narimanov.2}) and obtained the formula that accounts its influence on the level spacing statistics. The Podolskyi--Narimanov distribution \cite{Narimanov.3} needs two parameters: the value $\rho^{}_1$ same as for Berry--Robnik distribution, and the mean amplitude (normed to the same scale as the levels $\{\tilde E_\perp\}$) for the tunneling transitions $V^{}_{RC}$\,:
\begin{equation}\label{Pod.Nariman}
p^{}_{PN}(s) = \exp \left( -\rho^{}_1 s \right)
\left\{ \rho_1^2 \, \mathrm{erfc} \left( \frac{\sqrt{\pi}}{2} \rho^{}_2 s \right) F\left( \frac{s}{V^2_{RC}}\right) + \left( 2\rho^{}_1 \rho^{}_2 F\left( \frac{s}{V^{}_{RC}}\right) + \frac{\pi}{2} \rho_2^3 s \right) \exp \left( -\frac{\pi}{4} \rho_2^2 s^2 \right) \right\},
\end{equation}
where
\begin{equation*}
F(x) = 1 - \frac{1 - x \sqrt{\pi/2}}{\exp(x) - x} \,.
\end{equation*}

The aim of the present paper is to test the Podolskyi--Narimanov theory prediction using the set of the transverse motion energy levels of the high energy positrons channeling near $[100]$ direction of the silicon crystal. The motion in this case is characterised by the co-existence of regular and chaotic domains in the phase space as well as the substantial CAT rate, as it was demonstrated in \cite{RREPS.23.1}.

\section{Results and discussion}

\begin{figure}[htbp]
\vspace*{-2mm}\centering
\includegraphics[width=0.42\textwidth]{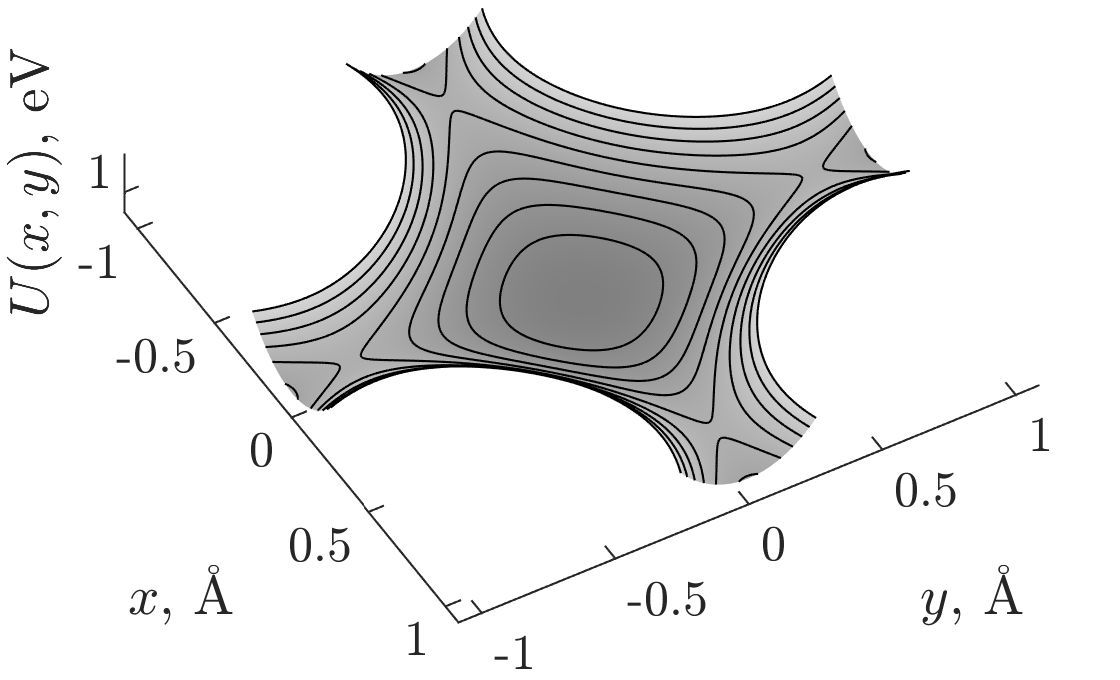} \ \ \ \ \ 
\includegraphics[width=0.38\textwidth]{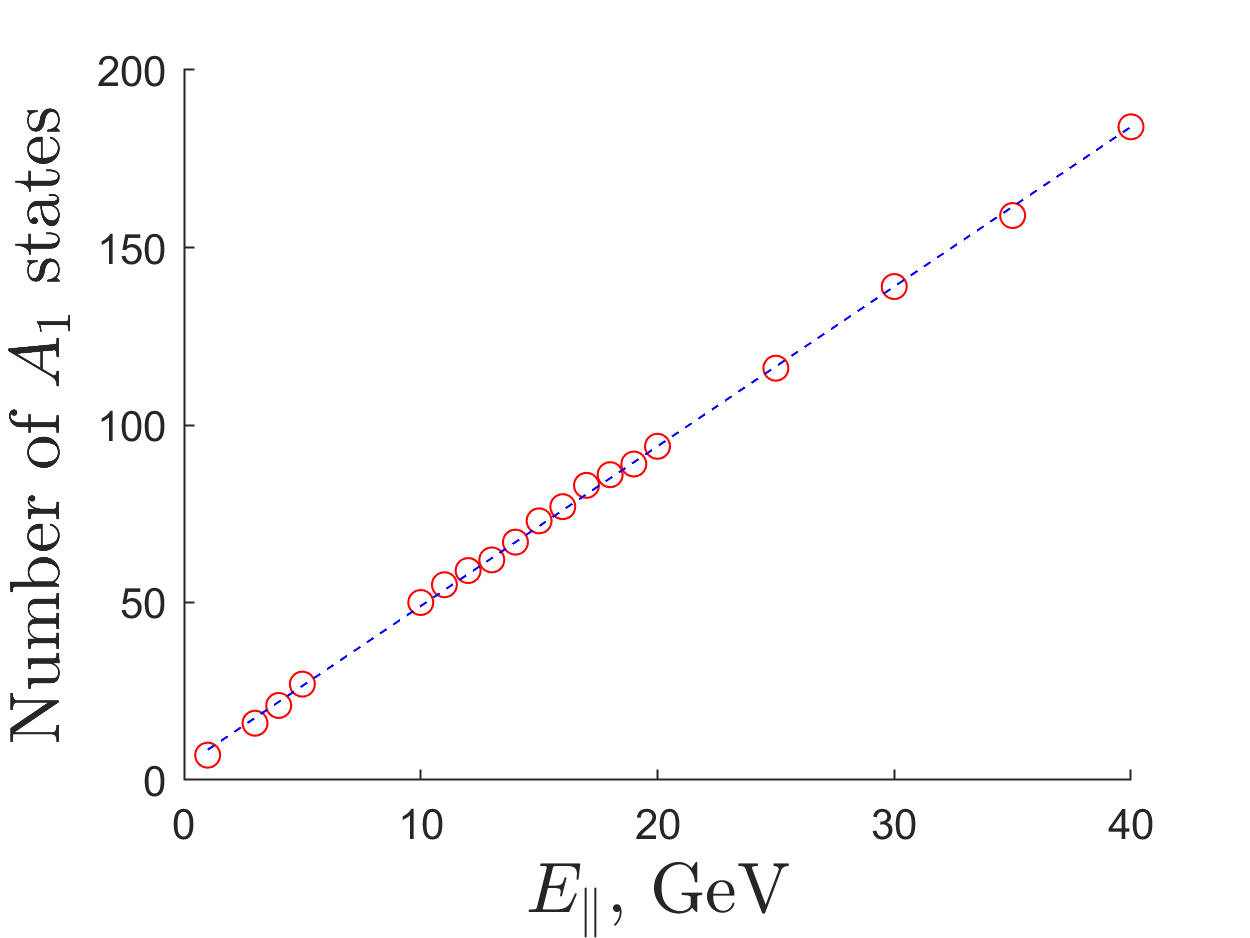}
\vspace*{-2mm}\caption{\label{potential.pic} Potential energy of the positron channeling near $[100]$ direction of the silicon crystal (\emph{left}) and the total number of the stationary states of $A_1$ type in this potential well as a function of $E_\parallel$ (\emph{right}).}
\end{figure}

The particle's transversal motion in the atomic string continuous potential is described by the two-dimensional Schr\"odinger equation with the parameter $E_{\parallel} / c^2$ (where $E_{\parallel} = (m^2 c^4 + p_{\parallel}^2 c^2)^{1/2}$) playing the role of the particle mass \cite{AhSh}. The quantum chaos manifestations takes the place and are studied in the semiclassical range of parameters, where the density of energy levels is high (and the classical orbits could be matched to the wave functions of the stationary states). The total number of the levels in the potential well (Figure \ref{potential.pic}, \emph{left}) grows with $E_\parallel$ (linearly in the 2-dimensional case, see Figure \ref{potential.pic}, \emph{right}) as it is predicted from semiclassical arguments \cite{AhSh}. Hence we have studied the level statistics for the positrons of energies high enough, from 20 to 40 GeV. The energy levels as well as the eigenfunctions corresponding to them are found numerically using the so-called spectral method \cite{3, Dabagov3, NIMB.2013}. 

The potential well where this transverse motion takes the place is formed by the repulsive uniform potentials of the four neighboing atomic strings $[100]$ of the silicon crystal (Figure \ref{potential.pic}, \emph{left}). This well possesses the symmetry of the square thus the stationary states of the channeling positrons can be classified using the irreducible representations of the group $C_{4v}$: four one-dimensional ones $A_1$, $A_2$, $B_1$, $B_2$ and one two-dimensional $E$ \cite{group, LL3}. So, the eigenstates of the first four types are non-degenerated while the last type ones are twice degenerated and are not of our interest. 

\begin{figure}[htbp]
\vspace*{-3mm}\centering
\includegraphics[width=0.49\textwidth]{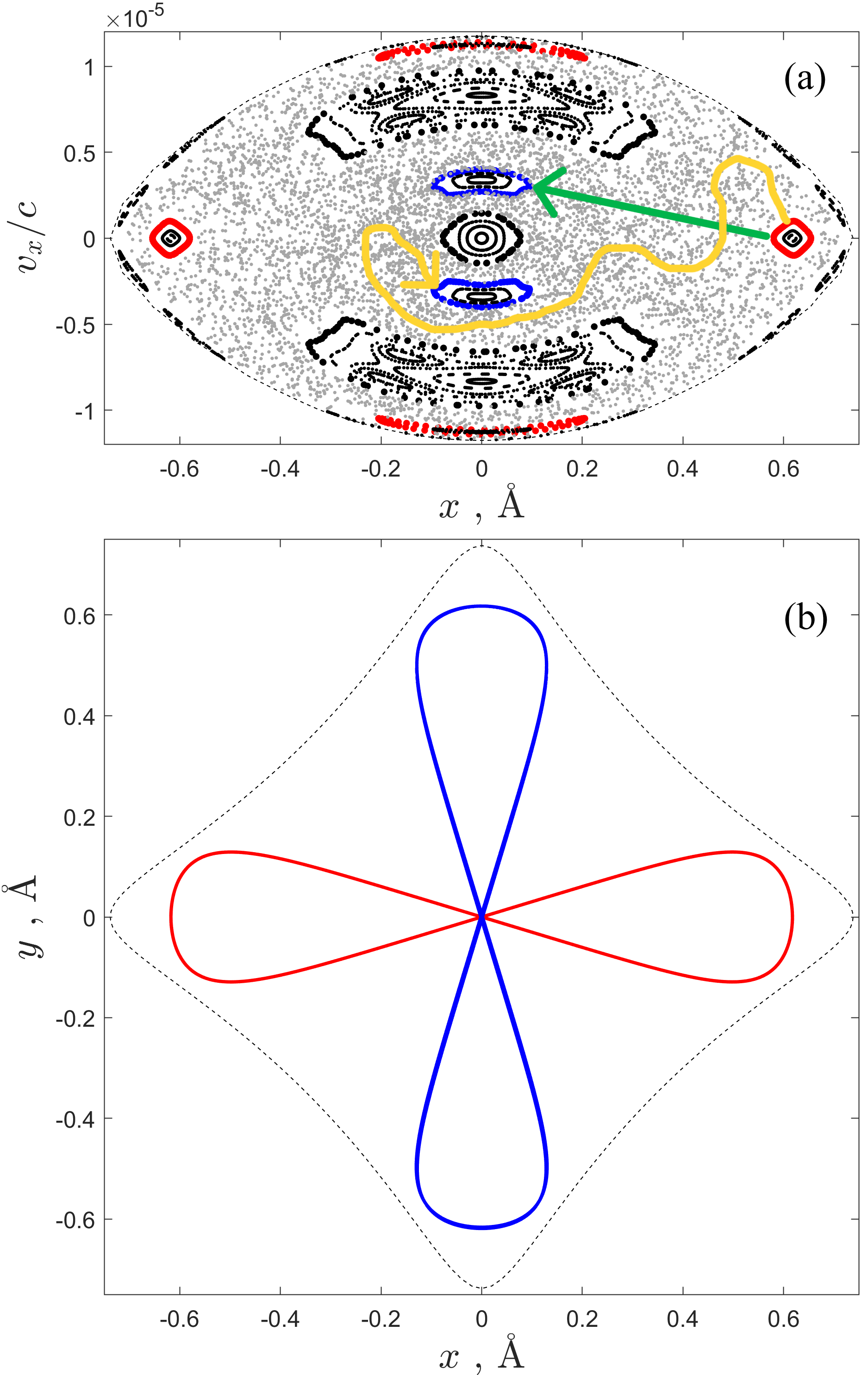} \ \ 
\includegraphics[width=0.49\textwidth]{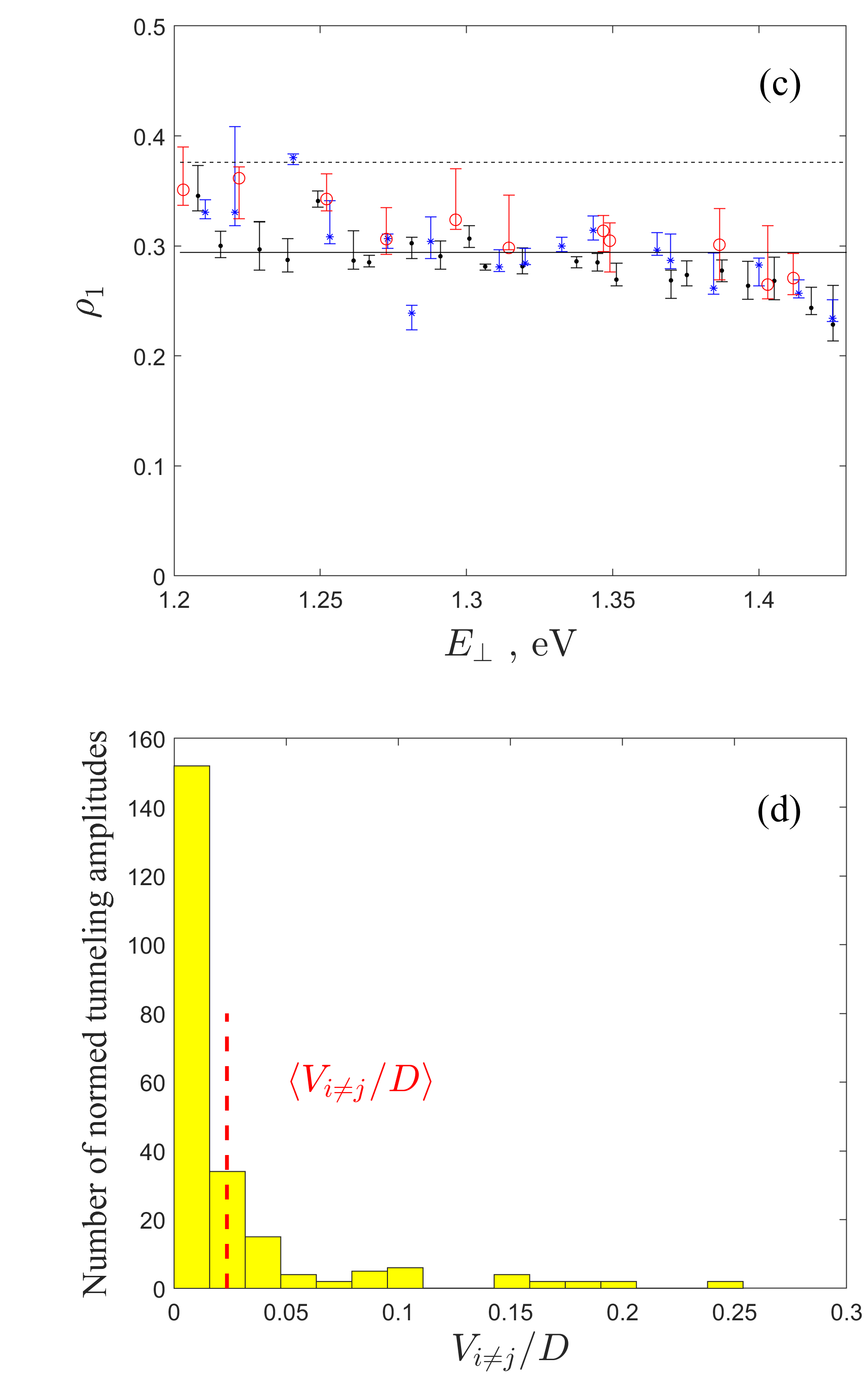} 
\vspace*{-9mm}\caption{\label{density} (\emph{a}): Poincar\'e section for the transverse motion at $E_\perp = 1.3865$ eV for $E_\parallel = 20$ GeV positrons. Green arrow schematically indicates the direct tunneling process between two regular motion domains, yellow arrow indicates CAT process. (\emph{b}): Two similar, but dynamically isolated periodic trajectories related to the regular motion domains marked in the panel (\emph{a}) by the corresponding color. (\emph{c}): Relative contribution of the regular dynamics domains in the phase space to the semiclassical mean density of states (\ref{density.via.Berry.Robnik.general}) calculated for 20 (red), 30 (blue), and 40 GeV positrons (black). (\emph{d}): Distribution of the absolute values of normed transition amplitudes.}
\end{figure}

The level spacing statistics has been studied on the transverse motion energy range
\begin{equation}\label{interval}
    1.2 \leq E_\perp \leq 1.43 \ \mbox{eV}
\end{equation}
(marked by the yellow band in Figure 4 ({\it left}) in \cite{RREPS.23.1}) where the relative contribution $\rho^{}_1$ from the regular domains in the phase space  (Figure \ref{density} (\emph{a})) to the mean level density (\ref{density.via.Berry.Robnik.general}) is approximately constant (see Figure \ref{density} (\emph{c})) as it is needed for  Berry--Robnik (\ref{Berry.Rob}) and Podolskyi--Narimanov (\ref{Pod.Nariman}) distributions. The method of obtaining the points in Figure \ref{density} (\emph{c}) is described in \cite{JINST.2019}. The weighted average value of $\rho_1$ on this range is
\begin{equation}\label{mean.rho.1}
    \rho^{}_1 = 0.294^{+0.020}_{-0.011} \,.
\end{equation}
The uncertainty of the estimate is due to the complexity of accounting the contribution of small domains of regular dynamics in the phase space of the channeled particle.

The smallness of the range (\ref{interval}) and hence the approximately constant mean level density $\rho (E_\perp)$ on it permit us to avoid the unfolding procedure and simply norm each level set to the unit mean level density. Note that each level set of the given $E_\parallel$ and the given symmetry class has to be normed independently,
\begin{equation}\label{renorming.1}
\{\tilde E_\perp\}_k = \{E_\perp\}_k / D_k 
\end{equation}
(where $D_k$ is the mean level spacing of the $k$-th original level set), and only after that the level spacings for the each set have to be collected to the array for the statistical analisys.

Podolskyi--Narimanov theory predicts that the levels interact via quantum mechanical tunelling of the semiclassical wave functions between classically isolated domains in the phase space, and the tunneling between two regular domains can occur in two ways \cite{Narimanov.1, Narimanov.2}, schematically indicated by the green and yellow arrows in Figure \ref{density} (\emph{a}). Direct tunneling is a low-probable first-order process described by a dimensionless (as well as $s$) amplitude with a characteristic value $V^{}_{RR} \ll 1$. Along with this, the second order CAT process is possible, in which the particle first tunnels only outside its own regular domain, and this process is described by the $V^{}_{RC}$ constant. The particle picked up by a chaotic flow can find itself near the boundary of another regular domain, where it tunnels inwards with the same amplitude $V_{RC}$. Thus, this is a second-order process, the resulting amplitude of which is $V_{RC}^2 \sim V^{}_{RR} \ll 1$ \cite{Narimanov.3}.

The technique used for estimating the amplitudes of these tunneling transitions $V_{ij}$ is described in \cite{RREPS.23.1}. The found amplitudes also need to be normalized to the average inter-level distance (Figure \ref{density} (\emph{d})). However, since the tunnel transitions happen between the superpositions of the states belonging to the definite symmetry class ($A_1$, $A_2$, $B_1$, $B_2$) \cite{RREPS.23.1}, the weighted average from all four $D_k$ values for the given $E_\parallel$ should be used for normalization. This value is $D = 0.0099$ for $E_\parallel = 20$ GeV, $D = 0.0065$ for $E_\parallel = 30$ GeV and $D = 0.0050$ for $E_\parallel = 40$ GeV. According to the pointed above, the parameter $V_{RC}$ in the Podolskyi--Narimanov distribution (\ref{Pod.Nariman}) should be put equal to the square root of the average normalized amplitudes:
\begin{equation}\label{VRC.1}
V^{}_{RC} = \sqrt{ \left< V_{ij} / D \right>}\,.
\end{equation}
For the set of arrays of energy levels in the range (\ref{interval}), this value is
\begin{equation}\label{VRC.2}
V^{}_{RC} = 0.1536 \,.
\end{equation}

\begin{figure}[htbp]
\vspace*{-2mm}\centering
\includegraphics[width=0.49\textwidth]{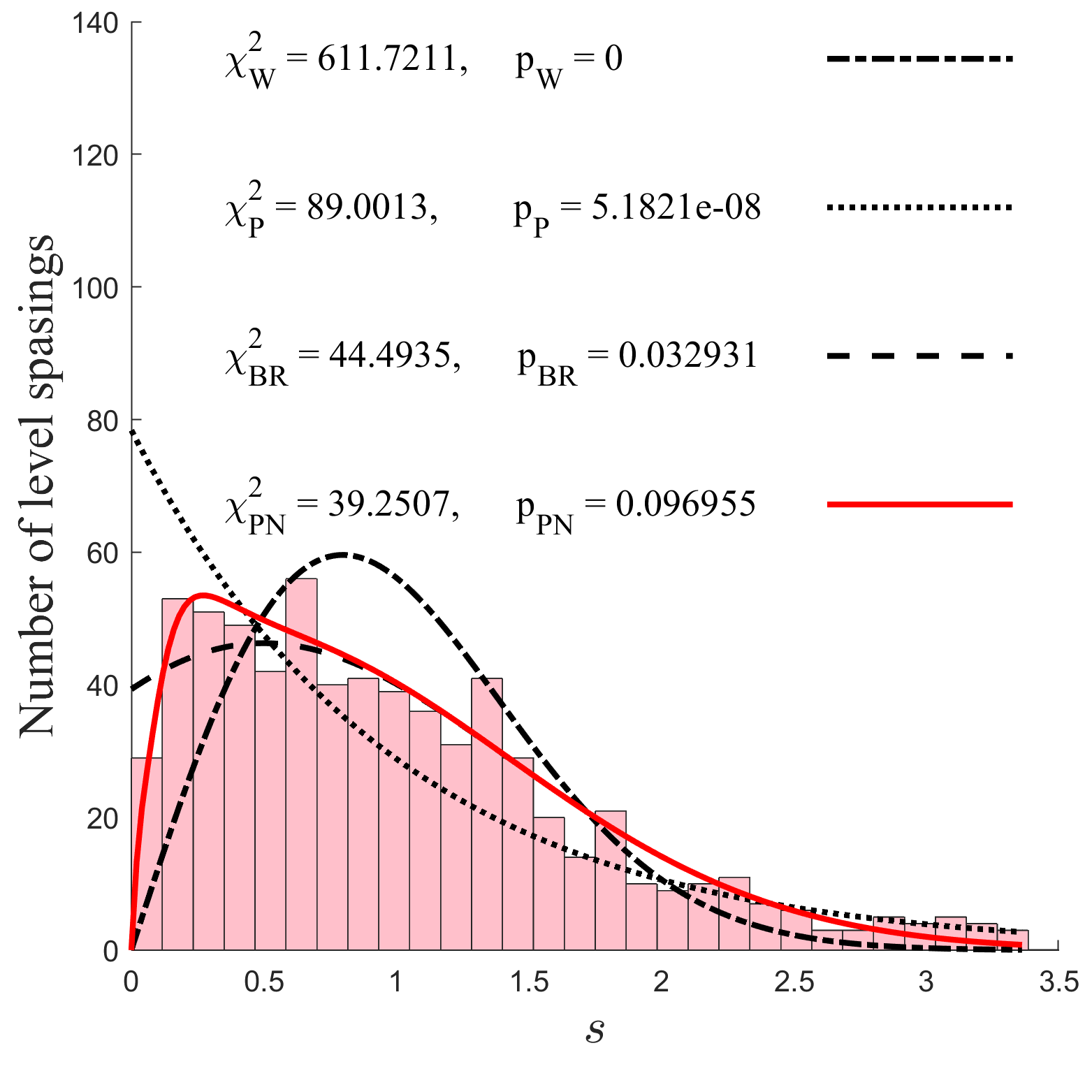} \ \ 
\includegraphics[width=0.49\textwidth]{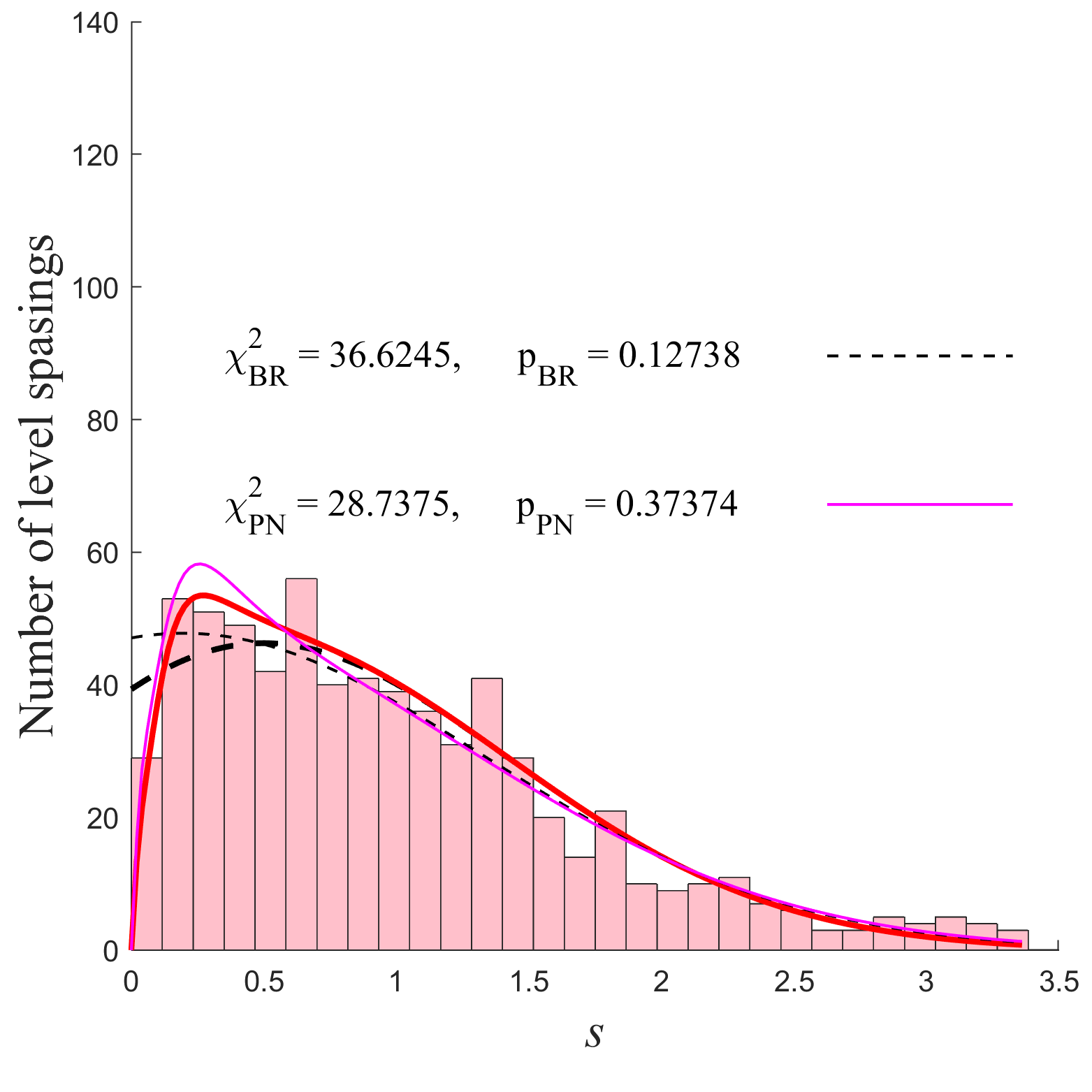}
\vspace*{-12mm}\caption{\label{hystogram} \emph{Left:} Level spacing distribution for the 20, 25, 30, 35 and 40 GeV positrons and the theoretical curves that correspond Wigner (dash-dotted), Poisson (dotted), Berry--Robnik (dashed) and Podolskiy--Narimanov (solid line) distributions, as well as $\chi^2$ values and the corresponding $p$-values (the parameters (\ref{mean.rho.1}) and (\ref{VRC.2}) are used). \emph{Right:} Berry--Robnik and Podolskiy--Narimanov distributions with the same parameters as in the left panel (thick lines), and with the parameters found according to the maximal likelihood criterion (thin lines).}
\end{figure}

The level spacing distribution in the range (\ref{interval}) for the positrons with $E_\parallel$ of 20, 25, 30, 35 and 40 GeV is given as a histogram in Figure \ref{hystogram} (\emph{left}). The curves corresponding to the predictions (\ref{Poisson})--(\ref{Pod.Nariman}) are superimposed on the histogram. For these four distributions, the values of $\chi^2$ and the corresponding $p$-values for 29 degrees of freedom are also determined. One can see that the distribution
of level spacings agrees well with the prediction of Podolskiy--Narimanov theory \cite{Narimanov.3}.

It is interesting also to plot the distributions (\ref{Berry.Rob}) and (\ref{Pod.Nariman}) with the free parameters fitted according to the maximal likelihood criterion. The corresponding curves are shown in Figure \ref{hystogram} (\emph{right}) by thin lines. For the Berry--Robnik distribution, the result of fitting is as follows:
\begin{equation}\label{mean.rho.2.fit}
    \left(\rho^{}_1\right)^{fit}_{BR} = 0.363 \,.
\end{equation}
In this case, the $\chi^2$ value and the corresponding $p$-value for 28 degrees of freedom (that takes into account the presence of one fitting parameter) are as follows:
\begin{equation}\label{chi2.fit.1}
    \chi^2_{BR} = 36.6245\,, \quad p^{}_{BR} = 0.12738 \,.
\end{equation}

The fitting results for the Podolskiy--Narimanov distribution are as follows:
\begin{equation}\label{mean.rho.3.fit}
    \left(\rho^{}_1\right)^{fit}_{PN} = 0.385 \,, \quad  \left(V^{}_{RC}\right)^{fit}_{PN} = 0.182 \,.
\end{equation}
The $\chi^2$ value and the corresponding $p$-value for 27 degrees of freedom (due to the presence of two fitting parameters) are as follows:
\begin{equation}\label{chi2.fit.2}
    \chi^2_{PN} = 28.7375\,, \quad p^{}_{PN} = 0.37374 \,.
\end{equation}
The similarity of the curves corresponding to the actual values of the system parameters and the curves found by the maximum likelihood criterion also indicates good agreement between the distribution of inter-level distances in the considered system and the predictions of quantum chaos theory.

\section{Conclusion}

The energies of the transverse motion stationary states for the positrons with the energies $E_\parallel = 20$, 25, 30, 35, and 40 GeV channeling in the [100] direction of the silicon crystal are found numerically. Through these stationary states the groups of them, which could be interpreted as an interaction result of the quasi-classical states, corresponding to particle localization in the dynamically separated regions of the phase space, are identified. The matrix element values for the transitions between these quasi-classical states are found. Earlier it was demonstrated that the values of these matrix elements are in agreement with the concept of chaos-assisted tunneling (CAT).

The statistical analysis of the inter-level distances array done in this paper shows that the Podolskiy--Narimanov distribution describes this array features better than the Berry--Robnik distribution, which not accounts the CAT effect. The distribution parameter values from the analysis of the system dynamics and from the most likelihood fit are similar that also points to a good agreement between the theory predictions and the numerically simulated distribution.

The difficulty in observation of the transversal motion quantization consists in the energy levels broadening due to dechanneling. However, the channeling of the positively charged particles appears relatively stable. For 50 GeV positrons the dechanneling length according to (B.5) of \cite{Korol} is $L_d \sim 1.5$ cm. The corresponding energy levels width is $\Gamma_\perp = c\hbar/L_d \sim 10^{-5}$ eV and less than level spacing according to our numerical calculations. So, the possibility to observe the influence of the discussed peculiarities on observable data exists in principle, however requires further study.

\end{document}